\documentclass[letter]{jpsj3} 
\usepackage{txfonts}
\usepackage{graphicx,amsmath,xspace}

\newcommand{\alphaI}{{$\alpha$-(BEDT-TTF)$_2$I$_3$}\xspace}
\newcommand{\be}{\begin{equation}}
\newcommand{\ee}{\end{equation}}
\newcommand{\bea}{\begin{eqnarray}}
\newcommand{\eea}{\end{eqnarray}}

\title{Spin-Ordered States in Multilayer Massless Dirac Fermion Systems}

\author{\name{Kenji Kubo}
  \thanks{E-mail: kubo.kenji.22x@st.kyoto-u.ac.jp}
  and 
  \name{Takao Morinari}
}

\inst{Graduate School of Human and Environmental Studies, Kyoto
  University, Kyoto 606-8501, Japan}

\abst{We investigate the spin-ordered states 
in multilayer massless Dirac fermion systems 
under magnetic fields, in which 
the intralayer interaction is ferromagnetic owing to the exchange interaction,
while the interlayer interaction is antiferromagnetic arising from
the interlayer hopping and the on-site Coulomb repulsion.
The possible spin-ordered states are examined within the mean field theory,
and we apply it to $\alpha$-(BEDT-TTF)$_2$I$_3$,
which is a multilayer massless Dirac fermion system under pressure.
In the weak interlayer coupling regime the system exhibits 
a ferromagnetically spin-ordered state with the effective Zeeman 
$g$-factor {\it less} than two contrasting to that observed 
in the single-layer graphene.
}

\usepackage{amsmath}
\begin{document}
\maketitle


A multilayer organic conductor 
\alphaI [BEDT-TTF=bis(ethylenedithio)tetrathiafulvalene] \cite{Bender1984} 
has attracted a great deal of attraction 
since it was found that the energy dispersion is linear under
pressure\cite{Kobayashi2004,Katayama2006,Kobayashi2007}. 
$\alpha$-(BEDT-TTF)$_2$I$_3$ has a layered structure, in which
conducting layers of BEDT-TTF molecules and insulating layers of I$_3$
anions stack alternatively. 
Below 135 K, charge ordering with a stripe pattern
makes $\alpha$-(BEDT-TTF)$_2$I$_3$ an insulating phase 
under ambient pressure.
\cite{Kino1995,Seo2000,Takano2001,Wojciechowski2003} 
For pressures higher than 1.5GPa,
the charge ordering transition is suppressed
and the system becomes metallic even at low temperature.\cite{Tajima2000}
The resistivity is almost temperature independent 
while the Hall coefficient shows 
strong temperature dependence.
\cite{Tajima2000}

Using the tight-binding model with 
the transfer integrals obtained by an X-ray diffraction 
experiment \cite{Kondo2005},
Kobayashi and cowoekers 
calculated the energy dispersion of 
$\alpha$-(BEDT-TTF)$_2$I$_3$.\cite{Kobayashi2004,Katayama2006}
They found that the band structure near the Fermi energy 
is described by a tilted 
and anisotropic Dirac cone, which was supported by 
the first principles calculation.\cite{Ishibashi2006,Kino2006}
The presence of Dirac fermions is clearly demonstrated in the interlayer
magnetoresistance measurement \cite{Tajima2009}
where the zero energy Landau level of Dirac fermions leads
to negative magnetoresistance.\cite{Osada2008} 

In graphene,\cite{Novoselov2004} which is 
a well-established Dirac fermion system, 
the existence of the Dirac fermion spectrum was
clearly demonstrated by the observation 
of the half-integer quantum Hall effect.
\cite{Novoselov2005,Zhang2005}
Under a high magnetic field,
lifting of spin degeneracy is observed experimentally.\cite{Zhang2006}
Nomura and MacDonald examined a criterion
for the occurrence of quantum Hall ferromagnet states 
at zero-temperature.
\cite{Nomura2006}
Under a magnetic field, the kinetic energy is quenched into the Landau levels
while the Landau level broadening plays the role of the band width.
In order to stabilize a quantum Hall ferromagnetic state, a cleaner 
system is plausible.
In this regard, we expect that a quantum Hall ferromagnetic state
is more stable in \alphaI than in graphene since the former 
is cleaner than the latter.\cite{Tajima2009}
We also expect that the multilayer structure of \alphaI 
should lead to symmetry broken states at finite temperature.

In this work, we investigate the possible spin-ordered state
in $\alpha$-(BEDT-TTF)$_2$I$_3$ within the mean field theory.
In our model, the intralayer ferromagnetic interaction
arises from the exchange interaction and the interlayer 
antiferromagnetic interaction arises from 
interlayer hopping and on-site Coulomb repulsion.
We also include the Zeeman energy term that plays an important role
in selecting a stable spin-ordered state.

We study multilayer massless Dirac fermion system
under a magnetic field.
In each layer, we consider a single component of Dirac fermions.
In general there are two Dirac points in the Brillouin zone.
Here, we assume that Dirac fermions are degenerate with respect to 
these valley degrees of freedom and we do not consider the possibility
of lifting valley degeneracy.
For the description of Dirac fermions in each layer,
we take the following Hamiltonian:
\begin{equation}
 H=\nu\begin{pmatrix}
       0&\left(p_x+eA_x\right)-i\left(p_y+eA_y\right)\\
       \left(p_x+eA_x\right)+i\left(p_y+eA_y\right)&0
      \end{pmatrix},\nonumber\\
\end{equation}
where $p_\alpha$ and $A_\alpha$ with $\alpha=x,y$ 
are momentum operators and the vector potential, respectively.
The velocity of Dirac fermions is denoted as $v$ and 
$-e$ is the electron charge.
We take the Landau gauge, $A_x=0$ and $A_y=Bx$, 
with $B$ being the applied magnetic field.
In \alphaI, the energy dispersion of Dirac fermions is described
by a tilted and anisotropic cone.\cite{Katayama2006}
However, under a magnetic field, tilting and anisotropy
introduce a renormalization of the velocity $v$.\cite{Morinari09,Goerbig08}
Thus, we assume that this renormalization effect
is already included in $v$. For the case of
$\alpha$-(BEDT-TTF)$_2$I$_3$, we take $v=10^7$ cm/s.\cite{Tajima2009}

Taking the plane wave form with the wave number $k$ 
in the $y$-direction,
the Landau level wave functions for Dirac fermions are given by
\begin{equation}
\psi_{n,k}\left(x,y \right)
= \frac{1}{\sqrt{L}}\exp \left( iky \right)
\phi_{n,k} \left( x \right),
\end{equation}
with $L$ being the system dimension.
The energy spectrum is $E_n = 
\operatorname{sgn}(n) \sqrt{2|n|}\hbar v/\ell_B$.
Here $n$, the Landau level index, 
is an integer and $\ell_B = \sqrt{\hbar/(eB)}$ is the magnetic
length.
The function $\phi_{n,k}(x)$ is given by
\begin{eqnarray}
\phi_{n,k}(x)
&=& \frac{C_n}{\sqrt{{\ell}_B}}
\left[
  \begin{pmatrix}-i\operatorname{sgn} (n)\\0
  \end{pmatrix}
h_{|n|-1}
\left(\frac{x}{{\ell}_B}+k{\ell}_B\right) 
+\begin{pmatrix}0\\1\end{pmatrix}
h_{|n|}\left(\frac{x}{{\ell}_B}+k{\ell}_B\right)
\right].
\end{eqnarray}
Here, $h_{|n|}(\xi)$ is the harmonic oscillator wave function
and the normalization constant $C_n$ is $C_0=1$ 
and $C_n=1/\sqrt{2}$ for $n\neq 0$.
In terms of these Landau level wave functions, 
the electron field operator is written as
\begin{equation}
 \hat{\psi}(x,y) = \sum_{n,k,\sigma}\psi_{n,k}(x,y)\hat{c}_{n,k,\sigma},
\end{equation}
where $\hat{c}_{n,k,\sigma}$ is 
the annihilation operator of Dirac fermions with 
the Landau level index $n$, the wave number $k$, and spin $\sigma$.
The density operator is defined by
$\hat{\rho}({\bf r})=\hat{\psi}^\dagger(x,y)\hat{\psi}(x,y)$.
The Fourier transform of $\hat{\rho}({\bf r})$ is
\begin{equation}
 \hat{\rho}_{\bf q}
=\int d^2{\bf r}\exp(-i{\bf q\cdot r})\hat{\rho}({\bf r})
=\sum_{n,n',k,\sigma}
F_{n,n',k}^{\bf q}
c^\dagger_{n,k,\sigma}c_{n',k+q_y,\sigma},
\end{equation}
where $F_{n,n',k}^{\bf q}$ is the Landau level form factor.
Using the density operator,
the Coulomb interaction is described as
\begin{equation}
 V_C=\frac{1}{2L^2} \sum_{\bf q}V_{\bf q}
 \hat{\rho}_{\bf q}
 \hat{\rho}_{\bf -q},
\end{equation}
where $V_{\bf q}=e^2/(2\epsilon q)$ 
with $\epsilon$ being the dielectric constant.

Now we introduce the mean field approximation for the exchange 
interaction
\cite{AndoUemura1974}:
\begin{eqnarray}
 V^{MF}_C&=&-\frac{1}{2L^2}\sum_{\bf q}V_{\bf q}
\left(\sum F_{n_2,n_2,k}^{\bf
	 q}F_{n_1,n'_2,k+q_y}^{\bf -q}\left<c_{n_2,k+q_y,\sigma}^\dagger
   c_{n_2,k+q_y,\sigma}\right>c_{n_1,k,\sigma}^\dagger
   c_{n'_2,k,\sigma}\right. \nonumber \\
& & + \left. \sum F_{n_1,n_2,k}^{\bf
     q}F_{n'_1,n_1,k+q_y}^{\bf -q}\left<c_{n_1,k,\sigma}^\dagger
   c_{n_1,k,\sigma}\right>c_{n'_1,k+q_y,\sigma}^\dagger
   c_{n_2,k+q_y,\sigma}\right).
\end{eqnarray} 
In the following, we consider \alphaI, and we assume that the
Fermi energy is at the Dirac point.
In this case, the zero energy Landau level, the presence of which
is a characteristic feature of Dirac fermions,
is at the Fermi energy. 
  We may consider only the zero energy Landau
  level since the $n=1$ Landau level energy, 
  $E_1 \simeq 10\sqrt{B}$
  is large enough compared with the Landau level width
  at low temperatures.\cite{Morinari10a}
  The Landau level mixing is important at high temperatures.
  For instance, the Landau level mixing is not negligible 
  for $T>10$ K at $B=1$ T.
  However, we are interested in low temperature behaviors
  and we do not consider the Landau level mixing.

The mean field Hamiltonian for the zero energy Landau level is 
\begin{equation}
  H^{MF}_0=\sum_{k,\sigma}\left[\varepsilon_k
    -\frac{1}{L^2} \sum_{\bf q}V_{\bf q}
    \exp\left(-\frac{q^2{\ell}_B^2}{2}\right)\left<c_{k+q_y,\sigma}^\dagger
    c_{k+q_y,\sigma}\right>\right]c_{k,\sigma}^\dagger c_{k,\sigma},
\label{eq_H_MF}
\end{equation}
where we have used 
$F_{0,0,k}^{\bf q}F_{0,0,k+q_y}^{\bf -q}
=\exp\left(-q^2{\ell}_B^2/2\right)$.
In eq.~(\ref{eq_H_MF}) we introduce 
$k$-dependent energy $\varepsilon_k$ in order to introduce
the broadening of the Landau level in the presence of disorder.
For simplicity, we assume that the density of states 
of the Landau level has the following form suggested from
the self-consistent Born approximation
 (SCBA)\cite{AndoUemura1974,Shon98}:
\begin{equation}
 D(\varepsilon)=\frac{4}{\pi \Gamma} 
 \sqrt{1-\left(\frac{2\varepsilon}{\Gamma} \right)^2},
\label{eq_dos}
\end{equation}
with $\Gamma$ being the Landau level width. 
  According to the SCBA, under high magnetic fields, 
  $\Gamma$ is proportional to
  $\sqrt{B}$.\cite{AndoUemura1974,Shon98} 
  However, for the reason we shall explain below, 
  we regard $\Gamma$ as a constant.
Within the mean field approximation, 
the electron self-energy satisfies
the following self-consistent equation:
\begin{equation}
 \Sigma_{\sigma}=-\frac{1}{L^2}\sum_{\bf q}V_{\bf q}
 \exp\left(-\frac{q^2{\ell}_B^2}{2}\right)
 \int^{\frac{\Gamma}{2}}_{-\frac{\Gamma}{2}}
 d\varepsilon D(\varepsilon)f(\varepsilon+\Sigma_{\sigma}),
\end{equation}
with $f(\varepsilon)$ being the Fermi distribution function.
In order to focus on the spin ordering, we ignore the $k$-dependence
of the self-energy.
The summation with respect to ${\bf q}$ is carried out exactly.
The spin-ordered state is found by solving
the following self-consistent equation:
\begin{eqnarray}
  m &\equiv & \Sigma_\uparrow-\Sigma_\downarrow \nonumber\\
  &=&C\int^{\frac{\Gamma}{2}}_{-\frac{\Gamma}{2}}d\varepsilon
  \sqrt{1-\left(\frac{2\varepsilon}{\Gamma}\right)^2}
  \left[f\left(\varepsilon - \frac{1}{2}m\right)
    -f\left(\varepsilon + \frac{1}{2}m\right)\right],
  \label{eq_sc}
\end{eqnarray}
where 
$C=\sqrt{8/\pi}\left({e^2}/{\epsilon {\ell}_B}\right)/{\Gamma}$.

\begin{figure}[htbp]
\begin{center}
  \includegraphics[width=1.0\columnwidth]{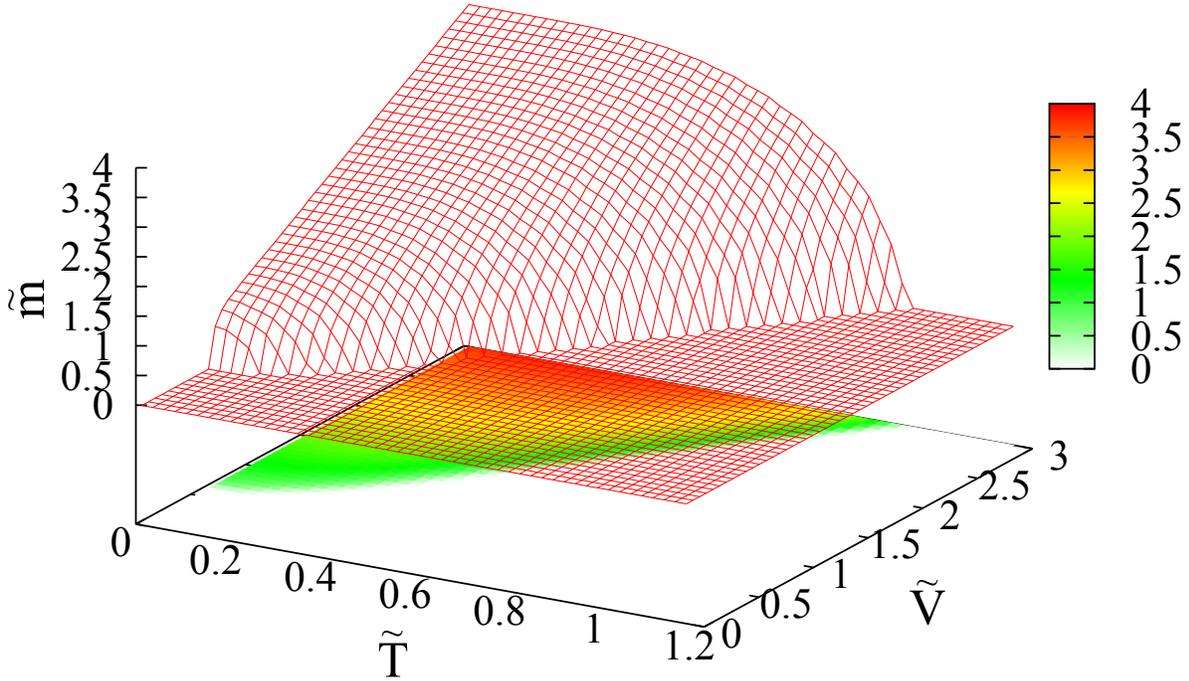}
 \caption{
   (Color online)
   Temperature, $\tilde{T}=T/\Gamma$, and Coulomb interaction, 
   $\tilde{V}=e^2/(\epsilon \ell_B \Gamma)$, 
   dependence of the order parameter, 
   $\tilde{m}=m/\Gamma$, for each layer of 
   \alphaI under pressure.
   We assume $\epsilon=190$ F/m for the dielectric constant, which is inferred from the analysis of the interlayer 
   magnetoresistance experiment.
   \cite{Tajima2010prb}
 }
\label{fig_pd}
\end{center}
\end{figure}

At zero temperature, the condition for the quantum Hall ferromagnetic 
state is
$\Gamma<\sqrt{8/\pi}(e^2)/(\epsilon {\ell}_B)$.
This corresponds to the Stoner criterion for itinerant 
ferromagnetism in a metal.
Using the parameter $\epsilon=190$ F/m
\cite{Tajima2010prb}
for \alphaI, we find
$\Gamma < 5.4 \sqrt{B}$
with $\Gamma$ measured in units of kelvin and $B$ measured in units of
tesla.
When we consider that $\Gamma$ is proportional to $\sqrt{B}$
under high magnetic field, $\Gamma = \alpha\sqrt{B}$, 
where $\alpha$ is a constant. 
According to Tajima {\it et al}., \cite{Tajima2009} 
the Landau level width at $T=1$ K is about $1.2$ K.
Below 1 K, the interlayer magnetoresistance minimum exists
at the magnetic field $B$ that satisfies $2\mu_BB/\Gamma\simeq 1$.
We define this $B$ as $B_0$.
From the analysis of the experiment,\cite{Tajima2009} we find that
$\alpha\sqrt{B_0}\simeq1.2$ K and $\alpha=1.3$ at $T=1$ K. 
At lower temperatures, the parameter $\alpha$ appears to decrease.
From this estimation of $\Gamma$, we may conclude that
the Stoner criterion is satisfied. 
Meanwhile, above 1 K, the width of Landau levels is mainly determined
by the temperature and $\Gamma$ is not proportional to $\sqrt{B}$. 
We surmise that \alphaI is so clean that
the $\sqrt{B}$ dependence of $\Gamma$ is not discernable.
Therefore, here we take the elliptic density of state eq.~(\ref{eq_dos})
as a phenomenological formula, and take a constant value for $\Gamma$.
If we consider the Zeeman energy and take $\Gamma = 2\mu_BB$,
then we find that the Stoner criterion is satisfied for $B<16$T. 
Thus, we may neglect the effect of spin splitting for 
the intralayer spin-ordered states below $B=16$ T.
At finite temperature, we solve eq.~(\ref{eq_sc}) numerically.
The result is shown in Fig.~\ref{fig_pd}.
From the interlayer magnetoresistance experiment,\cite{Tajima2009}
it was estimated that $\Gamma \sim 1$ K for $T<1$ K.
Therefore, we may conclude that the quantum Hall ferromagnetic state
is stabilized within each layer in \alphaI from Fig.~\ref{fig_pd}.

The calculation above is easily extended to the $n\neq0$ Landau levels,
which is relevant for hole-doped \alphaI.\cite{Tajima2013}
The difference is just the numerical factor of the Landau level form factors.
However, there is not so much difference for the spin-ordering criterion
itself between the $n=0$ case and the $n\neq 0$ case.
The critical $\Gamma$ for the $n=1$ Landau level is 
given by that for the $n=0$ Landau level
multiplied by the factor $11/16$.
However, we expect that $\Gamma$ takes large values 
in $n\neq 0$ Landau levels.
Therefore, the quantum Hall ferromagnetic state may be
unstable for high Landau levels.


Now we consider the interlayer coupling effect.
When the condition for $m\neq 0$ is satisfied,
each layer is in the spin-polarized state.
Taking the $z$-axis for the direction of the spin polarization,
we define $\hat{S}_{\bf q}^j \equiv ({\bf \hat{S}}_{\bf q}^j)_z$
by the Fourier transform of 
$\sum_\sigma \sigma 
\hat{\psi}^\dagger_{j\sigma} (x,y)
\hat{\psi}_{j\sigma} (x,y)$,
with $j$ being the layer index.
A crucial difference between graphene and \alphaI is that
the strong electron correlation plays an important role in \alphaI.
In fact, the system is insulating owing to the strong electron correlation
under ambient pressure.
\cite{Kino1995,Seo2000,Takano2001,Wojciechowski2003} 
On-site Coulomb repulsion $U$ and interlayer hopping $t_{\perp}$ lead
to the antiferromagnetic interaction,
$J' = 4t_{\perp}^2/U$, between layers.
The Hartree term associated with the interlayer antiferromagnetic interaction
is
\begin{eqnarray}
    J'\sum_{j,{\bf q}}\hat{S}^j_{\bf q}\hat{S}_{\bf -q}^{j+1}
  &\simeq&J'\sum_{j,\sigma,\sigma',{\bf q},k,k'}
  \sigma\sigma'
  \exp\left(-q^2{\ell}_B^2/2\right) \nonumber\\
  & & \times \left(\langle
  c^\dagger_{j,k,\sigma}c_{j,k+q_y,\sigma}\rangle
  c^\dagger_{j+1,k',\sigma'}c_{j+1,k'+q_y,\sigma'}\right.
  \nonumber\\
  & & \left. + c^\dagger_{j,k,\sigma}c_{j,k+q_y,\sigma}\langle 
  c^\dagger_{j+1,k',\sigma'}c_{j+1,k'+q_y,\sigma'}\rangle\right).
\end{eqnarray}
Hereafter, we only consider the zero energy Landau level, and
we denote $\hat{c}_{0,k,j}$ as $\hat{c}_{k,j}$.
We define the order parameter $m_j$ for the $j$-th layer
as $m_j=\sum_{\sigma}\sigma
\langle c^\dagger_{j,k,\sigma}c_{j,k,\sigma}\rangle$,
which is assumed to be $k$ independent in accordance with
the approximation introduced above.
In terms of these order parameters, the Hartree term is rewritten as
\begin{equation}
  J'\sum_{j,j',k,\sigma}\sigma\left(m_j
  c^\dagger_{j',k,\sigma}c_{j',k,\sigma}+m_{j'}c^\dagger_{j,k,\sigma}c_{j,k,\sigma}\right).
\end{equation}
The interaction between spins in the $j$-th layer is
\begin{eqnarray}
\lefteqn{-\sum_{{\bf q},j,k,\sigma}V_{\bf q}
  \exp\left(-\frac{q^2{\ell}_B^2}{2}\right)
  \left<c_{j,k+q_y,\sigma}^\dagger
  c_{j,k+q_y,\sigma}\right>c^\dagger_{j,k,\sigma}c_{j,k,\sigma}
  } \nonumber \\
& & \simeq -J\sum_{j,k,\sigma}\left(m_j\sigma+\rho_j \right)
c^\dagger_{j,k,\sigma}c_{j,k,\sigma},
\end{eqnarray}
where $\rho_j$ is the number density of the $j$-th layer
and we defined $J=\sqrt{\pi/8}e^2/(\epsilon {\ell}_B)$ 
for the intralayer ferromagnetic interaction parameter.

Including the Zeeman energy term the system is reduced to
the following Ising model:
\begin{equation}
 H_{MF}=\sum_{j}\sum_{i=1}^N\left\{-Jm_j s_j^i+J'
\left(m_js_{j+1}^i+m_{j+1}s_j^i\right)-\mu_BBs_j^i\right\},
\end{equation}
where $s_j^i$ is the spin at the $i$-th site in the $j$-th layer. 
Note that the parameter $J$ depends on the applied magnetic field $B$.
Reflecting the fact that the interlayer coupling is antiferromagnetic,
the order parameter $m_j$ takes different values
for $j$ even and for $j$ odd.
We denote the former and the latter as $m$ and $m'$, respectively.
The self-consitent equation for $m$,$m'$ is
\be
  \begin{split}
    m=\tanh\left[\beta\left(Jm-J'm'+\mu_BB\right)\right],\\
    m'=\tanh\left[\beta\left(Jm'-J'm+\mu_BB\right)\right].
  \end{split}
  \label{eq_m}
\ee

\begin{figure}[htbp]
\begin{center}
  \includegraphics[width=1.0\columnwidth]{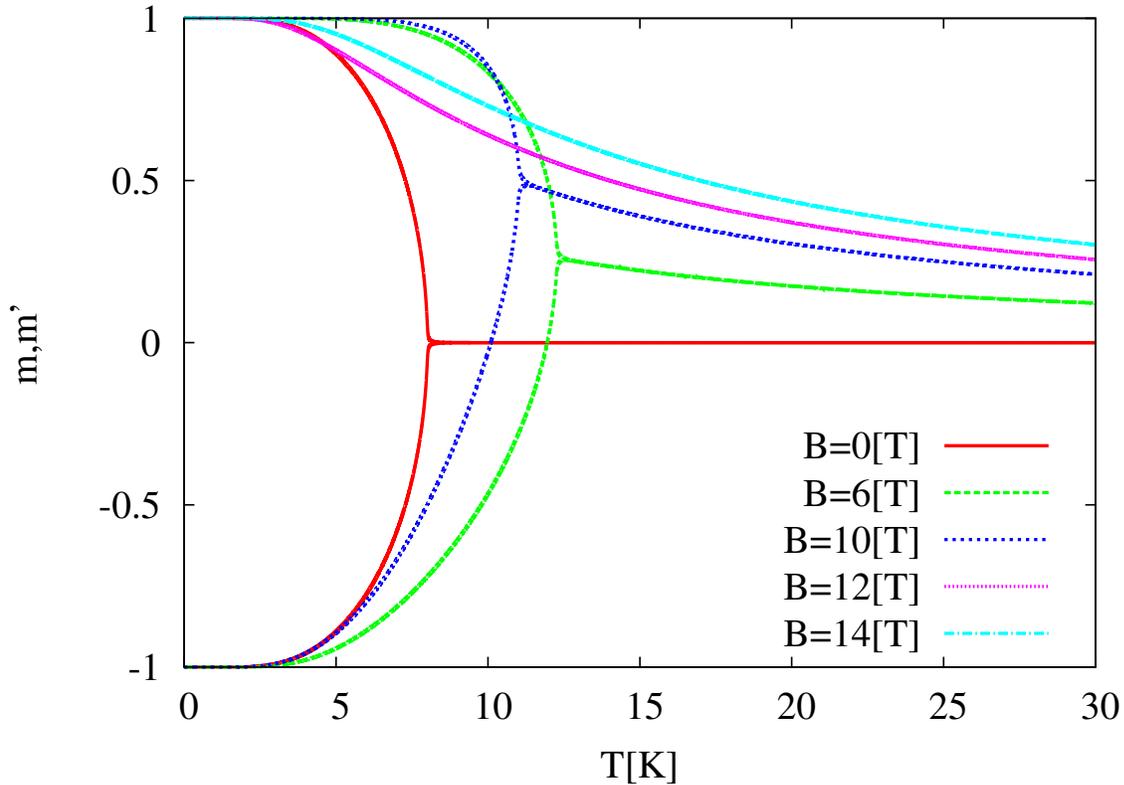}
\caption{
(Color online)
Temperature dependence of $m$ and $m'$ for different values of 
$B$ at $J'=8$ K.
The ferrimagnetic states can be realized 
in the weak magnetic field regime. 
In the strong magnetic field regime the ferromagnetically spin-ordered
state is stabilized.
}
\label{fig_op}
\end{center}
\end{figure}

\begin{figure}[htbp]
\begin{center}
  \includegraphics[width=1.0\columnwidth]{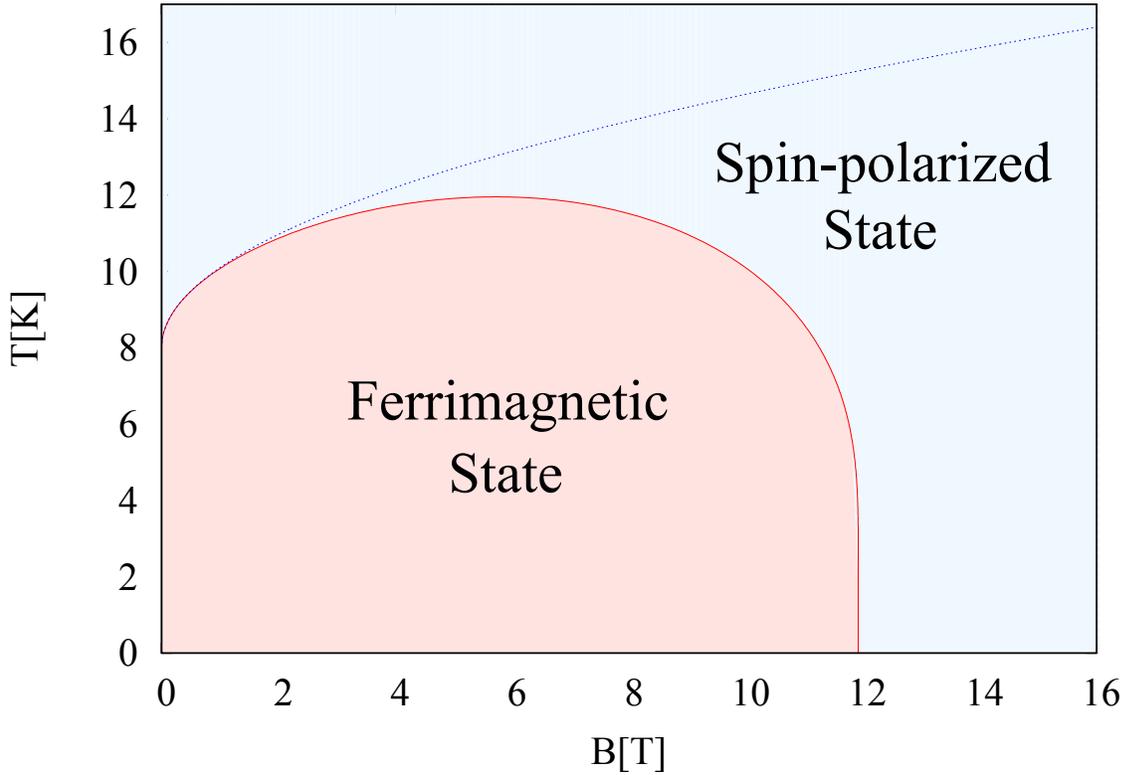}
\caption{
  (Color online)
  Phase diagram of spin-ordered states of the multilayer
 massless Dirac fermion system at $J'=8$ K.
 The critical temperature for
   the interlayer ferrimagnetic state is given by the solid
   line, eq.~(\ref{eq_Tc}). 
   The dotted line shows the critical temperature
   for the interlayer antiferromagnetic state in the absence of 
   the Zeeman energy.
}
\label{f5}
\label{fig_phase_diagram}
\end{center}
\end{figure}

\begin{figure}[htbp]
\begin{center}
  \includegraphics[width=1.0\columnwidth]{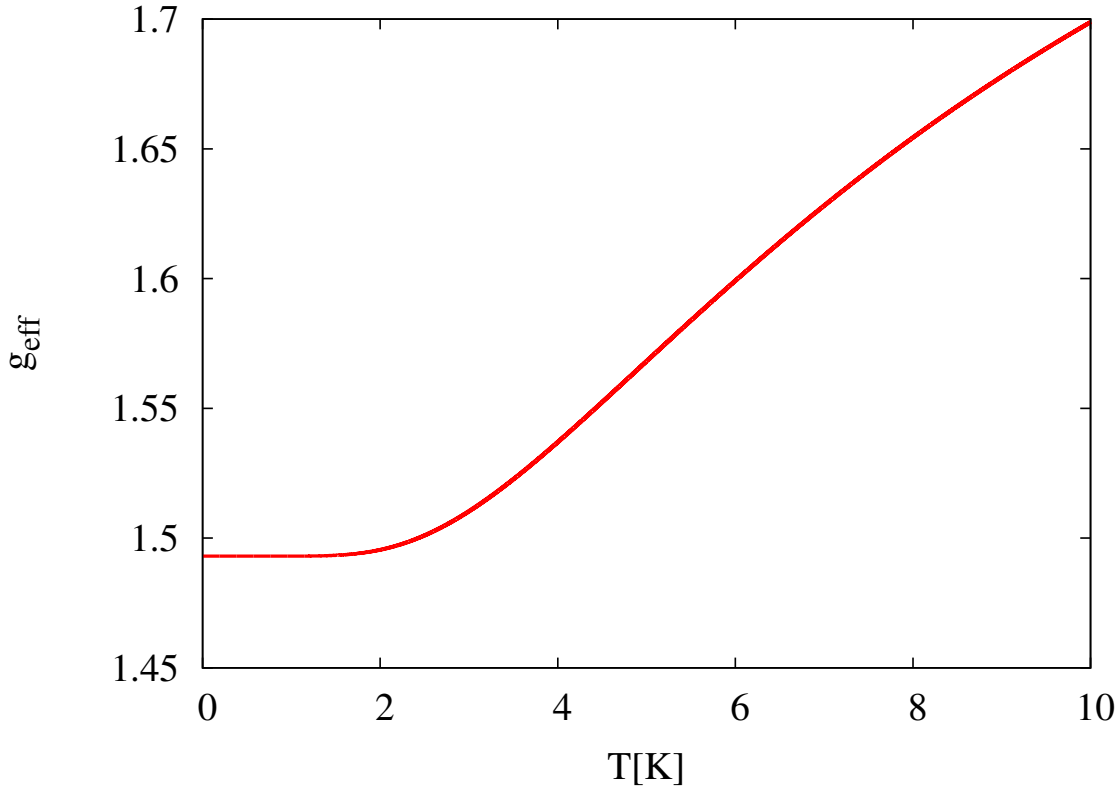}
\caption{
(Color online)
Temperature dependence of $g_{\rm eff}$.
The effective $g$-factor, $g_{\rm eff}$, is less than $g=2$
and decreases at low temperatures.
This behavior is observed when $2.1\sqrt{B}<J'<\mu_BB$ is satisfied.
}
\label{fig_geff}
\end{center}
\end{figure}

Here, $\mu_B$ is the Bohr magneton 
and $\beta = 1/(k_B T)$ with $k_B$ being the Boltzmann constant.
We solve this self-consistent equation numerically 
and obtained Fig.~\ref{fig_op} at $J'=8$ K.
Here, we assume a relatively large value for $J'$, 
which is the same order of magnitude as the interlayer hopping
estimated in a related organic compound.\cite{Jindo2006}
The parameter $J'$ can be smaller
depending on the ratio of the interlayer hopping
to the on-site Coulomb repulsion.
An antiferromagnetically spin-ordered state is possible
only when $B=0$ T. 
For $B>0$ T, the spins are in a ferrimagnetically ordered state
because of the Zeeman energy effect at low temperatures.  
The spin-polarized state is stabilized under high-magnetic fields
where the Zeeman energy is larger than the interlayer
antifferomagnetic interaction.
The critical temperature $T_c$ for the magnetically ordering 
transition is obtained by substituting $m'=0$ into eq.~(\ref{eq_m}):
\begin{equation}
 T_c=\frac{\mu_BB\left(J/J'+1\right)}{\tanh^{-1}\left(\mu_BB/J'\right)}.
 \label{eq_Tc}
\end{equation}
Note that $J$ depends on $B$.
The phase diagram is presented in Fig.~\ref{fig_phase_diagram}.
The system is ferrimagnetic for $T<T_c$ and
spin-polarized for $T>T_c$. 
When $J'<\mu_BB$, the antiferromagnetic interaction is irrelevant
and only the spin-polarized state is stabilized.
Even in the spin-polarized state, a unique feature appears
that is distinct from graphene.
We introduce the effective $g$-factor as
\begin{equation}
 g_{\rm eff}=g+\frac{2}{\mu_BB}\left(Jm-J'm'\right),
\end{equation}
with $g=2$ being the $g$-factor in the vacuum.
The temperature dependence of $g_{\rm eff}$ is shown in Fig.~\ref{fig_geff}.
Although the spins are ferromagnetically ordered, the mean fields
associated with the neighboring layers suppress the energy splitting
owing to the Zeeman energy because the interlayer coupling is 
antiferromagnetic.
As a consequence, $g_{\rm eff}$ is less than $g$.
This behavior is in sharp contrast to that of graphene where
the effective $g$-factor becomes larger than $g$.\cite{Young2012}
This temperature dependence is consistent with
the experiment in \alphaI.
\cite{Tajima2013private}
To conclude, we have examined the spin-ordered states
in multilayer massless Dirac fermion systems.
The exchange interaction leads to the ferromagnetic 
intralayer interaction while the strong electron correlation
and the interlayer hopping lead to the antiferromagnetic interlayer
interaction.
Within the mean field theory, we have determined the phase diagram
relevant for \alphaI.
When the Fermi energy is at the Dirac point,
the system exhibits the quantum Hall ferromagnetic state
for $\Gamma<5.4\sqrt{B}$.
The interlayer antiferromagnetic interaction leads to the 
ferrimagnetic state in the weak magnetic field
regime.
Even in a spin-polarized state, we expect an unusual
behavior of the effective Zeeman $g$-factor, which is qualitatively
consistent with the experiment.
\cite{Tajima2013private}

{\footnotesize \section*{Acknowledgements} 
We are thankful to Naoya Tajima for helpful discussions.
This work was financially supported in part by 
a Grant-in-Aid for Scientific Research (A) 
on ``Dirac Electrons in Solids'' (No. 24244053) and 
a Grant-in-Aid for Scientific Research (C) (No. 24540370) 
from The Ministry of Education, Culture, Sports, 
Science and Technology, Japan.
}

\bibliography{./}

\end{document}